\documentclass[adp,a4paper,fleqn,12pt]{w-art}
\usepackage{makeidx}
\usepackage{amsfonts}
\usepackage{graphicx}
\usepackage{amsmath}
\usepackage{amssymb}
\usepackage{times,cite,w-thm}

\begin{document}

\DOIsuffix{theDOIsuffix}

\Volume{XX} \Month{XX} \Year{XXXX}

\Receiveddate{XXXX} \Reviseddate{XXXX} \Accepteddate{XXXX}
\Dateposted{XXXX}


\title{On Dark Energy and Accelerated Reference
Frames}

\author[Ulhoa]{S.C. Ulhoa\inst{1,}
  \footnote{Corresponding author\quad
  E-mail:~\textsf{sc.ulhoa@gmail.com}}}
\address[\inst{1}]{University of Brasilia,\\
Campus Gama, DF, 72405-610, Brazil.}

\begin{abstract}
The paper is devoted to an explanation of the accelerated rate of
expansion of the Universe. Usually the acceleration of the Universe,
which is described by FRW metric, is due to dark energy. It is shown
that this effect may be considered as a manifestation of torsion
tensor for a flat Universe in the realm of Teleparallel gravity. An
observer with radial field velocity obey Hubble's Law. As a
consequence it is established that this is radial acceleration in a
flat Universe. In Eq. (\ref{24}) the acceleration is written in
terms of the deceleration parameter, the Hubble's constant and the
proper distance. This may be interpreted as an acceleration of the
Universe.
\end{abstract}


\maketitle

\section{Introduction}
\noindent

An accelerated expansion of the Universe has been verified by recent
observational data~\cite{Perlmutter:1998np,Riess:1998cb}. The
existence of dark energy is considered as an explanation of this
fact~\cite{jimenez:063005,2003astro.ph..7335S,Chen:2009bca,Tsujikawa:2010sc,Silvestri:2009hh},
even though the nature of such an energy remains obscure. The main
feature of this energy, that is almost 70\% of the content of the
Universe, is that it has a repulsive gravitational force.

There are several ansatz about the nature of the dark energy. One of
them associates cosmological constant to dark
energy~\cite{2003RvMP...75..559P,Feng:2000if,Kratochvil:2004gq}.
This explanation is popular since it behaves as a cosmological fluid
with  small constant energy density and negative pressure. In this
model the equation of state is $\omega=\frac{p}{\rho}=-1$, where p
is the pressure and $\rho$ is the fluid density. There is no time
evolution in such a model for $\omega$. Other models of dark energy
do have some time dependence.

A scalar field has been considered as the source of such an
energy~\cite{Sergijenko:2008md,Alimohammadi:2009yt,PhysRevD.61.083502,2005GReGr..37.1427S,2008mgm..conf..824M}.
This includes the well-known \textit{quintessence}
model\cite{1988NuPhB.302..668W}. In such models the negative
pressure arises when the potential energy becomes dominant in the
temporal evolution. Such a behavior is obtained if the kinetic term
is proportional to the derivative of the potential energy. A third
option is the so called \textit{f(R) theories}. The idea is to
modify the Einstein-Hilbert lagrangian density adding functions of
Ricci scalar which become important at late times and for small
scalar
curvature~\cite{PhysRevD.68.063510,Rador:2007wq,Nojiri:2003ni}. This
procedure modifies the Friedmann's equation with an extra
acceleration term.

Recently in the realm of Einstein-Cartan theory, it is suggested
that dark energy could arise from the coupling between torsion
tensor and fermions. A cosmological BCS mechanism is used to derive
the effective equations of motion which leads to an
acceleration~\cite{PhysRevD.81.043511}. Another approach, in the
same context, is given in \cite{ANDP:ANDP201000162}, where an
effective lagrangian density is used to derive an effective
cosmological constant.

In this paper a possible explanation is given for the accelerated
expansion of the Universe based on the teleparallel gravity. The
main advantage of this method is that it leads to an expression for
the acceleration naturally in terms of the torsion tensor. It does
not need ad-hoc modifications of the Einstein-Hilbert equations in
order to accommodate the concept of dark energy. In fact Schucking
showed that the curvature does not describe acceleration but simply
its gradient\cite{Schucking}.

The paper is organized as follows. In section 2, the teleparallel
gravity is introduced and an accelerated reference frame in
spacetime is defined. In section 3 the FRW metric is used to
describe the Universe. The acceleration of an observer in a galaxy
in a radial velocity field is calculated, considering the Earth as
stationary. In particular the flat Universe is analyzed. Section 4
has some concluding remarks.

\bigskip
Notation: space-time indices $\mu, \nu, ...$ and SO(3,1) indices $a,
b, ...$ run from 0 to 3. Time and space indices are indicated
according to $\mu=0,i,\;\;a=(0),(i)$. The flat, Minkowski space-time
metric tensor raises and lowers tetrad indices and is fixed by
$\eta_{ab}=e_{a\mu} e_{b\nu}g^{\mu\nu}= (-+++)$. The determinant of
the tetrad field is represented by $e=\det(e^a\,_\mu)$. It is used
$G=c=1$.

\section{Accelerated Reference Frames in Teleparallel Gravity}
\noindent

In General Relativity the metric tensor $g_{\mu\nu}$ plays the role
of a dynamical variable. This leads to the curvature tensor in terms
of the Christoffel symbols ${}^0\Gamma^\lambda_{\mu\nu}$ and the
corresponding torsion tensor vanishes. However, the field equations
in teleparallel gravity are constructed using the torsion tensor of
the Weitzenb\"ock space-time~\cite{hehl-1994}, defined as

\begin{equation}
T^{\lambda}\,_{\mu\nu}(e)=e_a\,^{\lambda}(\partial_\mu
e^{a}\,_{\nu}-\partial_\nu e^{a}\,_{\mu})\,. \label{0.0}
\end{equation}
The dynamical variable in the teleparallel gravity is the tetrad
field $e^{a}\,_{\nu}$. The curvature tensor vanishes identically. It
should be noted that $T^{\lambda}\,_{\mu\nu}$ is also the object of
anholonomity.

Let ${}^0\omega_{\mu ab}$ represent the torsion-free Levi-Civita
connection,

\begin{eqnarray}
^0\omega_{\mu ab}&=&-{1\over 2}e^c\,_\mu(
\Omega_{abc}-\Omega_{bac}-\Omega_{cab})\,,\label{0.001}
\end{eqnarray}

with

\begin{eqnarray} \Omega_{abc}&=&e_{a\nu}(e_b\,^\mu\partial_\mu
e_c\,^\nu-e_c\,^\mu\partial_\mu e_b\,^\nu)\,,\label{0.1}
\end{eqnarray}
The Levi-Civita connection is related to
${}^0\Gamma^\lambda_{\mu\nu}$ as

\begin{equation}
^0\Gamma^\lambda_{\mu\nu}=\Gamma^\lambda_{\mu\nu}+
e^{a\lambda}e^b\,_\nu\, {}^0\omega_{\mu ab}\,,\label{0.01}
\end{equation}
where $\Gamma^\lambda_{\mu\nu}$ the Weitzenb\"ock connection.

The following identity is relevant in the construction of the
teleparallel gravity,

\begin{equation}
^0\omega_{\mu ab}=-K_{\mu ab}\,, \label{0.2}
\end{equation}
where $K_{\mu ab}=\frac{1}{2}e_{a}\,^{\lambda}e_{b}\,^{\nu}
(T_{\lambda\mu\nu}+T_{\nu\lambda\mu}+T_{\mu\lambda\nu})$ is the
contorsion tensor of the Weitzenb\"ock connection.

Then the curvature scalar is given by

\begin{equation}
eR(^\circ\omega)\equiv -e\left({1\over 4}T^{abc}T_{abc}+{1\over
2}T^{abc}T_{bac}-T^aT_a\right) +2\partial_\mu(eT^\mu)\,.\label{0.3}
\end{equation}
with $T^a=T^b\,_b\,^a$. Both sides of (\ref{0.3}) are invariant
under Lorentz transformations. Dropping the divergence term the
Lagrangian density is

\begin{eqnarray}
{\cal L}(e_{a\mu})&=&-k\,e\,\left({1\over 4}T^{abc}T_{abc}+
{1\over 2} T^{abc}T_{bac} -T^aT_a\right)-{{\cal L}}_M\nonumber \\
&\equiv&-k\,e \Sigma^{abc}T_{abc} -{{\cal L}}_M \;,\label{0.4}
\end{eqnarray}
where $k=1/(16 \pi)$, ${{\cal L}}_M$ is the Lagrangian density for
the matter fields and $\Sigma^{abc}$ is defined by

\begin{equation}
\Sigma^{abc}={1\over 4} (T^{abc}+T^{bac}-T^{cab}) +{1\over 2}(
\eta^{ac}T^b-\eta^{ab}T^c)\;.\label{0.5}
\end{equation}

The field equation is obtained using a variational derivative with
respect to $e_{a\lambda}$

\begin{equation}
\partial_\nu(e\Sigma^{a\lambda\nu})={1\over {4k}}
e\, e^a\,_\mu( t^{\lambda \mu} + T^{\lambda \mu})\;,\label{0.6}
\end{equation}
where

\begin{equation}
t^{\lambda \mu}=k(4\Sigma^{bc\lambda}T_{bc}\,^\mu- g^{\lambda
\mu}\Sigma^{bcd}T_{bcd})\,,\label{0.7}
\end{equation}
and $e^a\,_\mu T^{\lambda \mu}=\frac{1}{e}\frac{\delta {{\cal
L}}_M}{\delta e_{a\lambda}}$ defines the energy-momentum tensor of
the matter field. It is possible to show that Eq. (\ref{0.6}) is
equivalent to the Einstein field equations.

Tetrad fields may be interpreted as reference
frames~\cite{Hehl2,Maluf:2007qq} adapted to a class of observers in
spacetime. If the world line $C$ of an observer in spacetime is
given by $x^\mu(\tau)$ and its velocity by
$u^\mu(\tau)=dx^\mu/d\tau$ where $\tau$ is the proper time, then the
observer's velocity is identified with $e_{(0)}\,^\mu$. Thus
$u^\mu(\tau)=e_{(0)}\,^\mu$ along $C$. The acceleration $a^\mu$ is
given by the derivative of $u^\mu$ along
$C$~\cite{Mashhoon:2003hv,Mashhoon:2002xs}, i. e.

\begin{equation}
a^\mu= {{Du^\mu}\over{d\tau}} ={{De_{(0)}\,^\mu}\over {d\tau}} =
u^\alpha \nabla_\alpha e_{(0)}\,^\mu\,, \label{1}
\end{equation}
where the covariant derivative is calculated using Christoffel
symbols which implies that the manifold preserves parallel
transport, with Weitzenb\"ock connection $a^\mu$ will vanish.

Following~\cite{Mashhoon:2002xs,Mashhoon:2003hv}, the absolute
derivative of $e_a\,^\mu$ is given by

\begin{equation}
{{D e_a\,^\mu} \over {d\tau}}=\phi_a\,^b\,e_b\,^\mu\,, \label{2}
\end{equation}
where $\phi_a\,^b$ is the acceleration tensor. The acceleration
tensor divides the acceleration of the whole frame along $C$. In
analogy with the Faraday tensor the identification $\phi_{ab}
\rightarrow ({\bf a}, {\bf \Omega})$, where ${\bf a}$ is the
translational acceleration ($\phi_{(0)(i)}=a_{(i)}$) and ${\bf
\Omega}$ is the frequency of rotation of the local spatial frame
with respect to a nonrotating (Fermi-Walker transported) frame,
proves to be plausible~\cite{Maluf:2007qq,Hehl2,Maluf:2009ey}.

In view of the orthogonality between tetrads it is possible to
isolate $\phi_a\,^b$ in relation (\ref{2}), i. e.

\begin{equation}
\phi_a\,^b= e^b\,_\mu {{D e_a\,^\mu} \over {d\tau}}= e^b\,_\mu
\,u^\lambda\nabla_\lambda e_a\,^\mu\,, \label{3}
\end{equation}
where $\nabla_\lambda e^b\,_\mu=\partial_\lambda e^b\,_\mu-
\,^0\Gamma^\sigma_{\lambda \mu} e^b\,_\sigma$. The meaning of such
tensor may be made clearer when the acceleration vector is written
in terms of the acceleration tensor. The projection of $a^\mu$ in
(\ref{1}) on a frame yields

\begin{equation}
a^b= e^b\,_\mu a^\mu=e^b\,_\mu u^\alpha \nabla_\alpha
e_{(0)}\,^\mu=\phi_{(0)}\,^b\,. \label{4}
\end{equation}
Therefore $a^\mu$ and $\phi_{(0)b}$ are not different accelerations
of the frame.

The expression of $a^\mu$ given by Eq. (\ref{1}) may be rewritten as

\begin{eqnarray}
a^\mu&=& u^\alpha \nabla_\alpha e_{(0)}\,^\mu =u^\alpha
\nabla_\alpha u^\mu = {{dx^\alpha}\over {d\tau}}\biggl( {{\partial
u^\mu}\over{\partial x^\alpha}}
+\,^0\Gamma^\mu_{\alpha\beta}u^\beta \biggr) \nonumber \\
&=&{{d^2 x^\mu}\over {d\tau^2}}+\,^0\Gamma^\mu_{\alpha\beta}
{{dx^\alpha}\over{d\tau}} {{dx^\beta}\over{d\tau}}\,. \label{5}
\end{eqnarray}
If $u^\mu=e_{(0)}\,^\mu$ represents a geodesic trajectory, then the
frame is in free fall with $a^\mu=0$ which means $\phi_{(0)(i)}=0$.
Therefore the nonvanishing values of $\phi_{(0)(i)}$ do represent
non-geodesic accelerations of the frame.

Due to the orthogonality of the tetrads, eq. (\ref{3}) is written as
$\phi_a\,^b= -u^\lambda e_a\,^\mu \nabla_\lambda e^b\,_\mu$, where
$\nabla_\lambda e^b\,_\mu=\partial_\lambda e^b\,_\mu- \,
^0\Gamma^\sigma_{\lambda \mu} e^b\,_\sigma$. Combining with Eq.
(\ref{0.01}), the components $\phi_a\,^b$ may be expressed as

\begin{equation}
\phi_a\,^b=e_{(0)}\,^\mu(\,\,^0\omega_\mu\,^b\,_a)\,. \label{6}
\end{equation}
Taking into account Eq. (\ref{0.2}), we get

\begin{equation}
\phi_{ab}={1\over 2} \lbrack T_{(0)ab}+T_{a(0)b}-T_{b(0)a}
\rbrack\,. \label{8}
\end{equation}
This is the acceleration
tensor~\cite{Mashhoon1990147,Mashhoon1990176} expressed in terms of
components of the torsion tensor of the Weitzenb\"ock
space-time~\cite{Maluf:2007qq}. The symmetric part of acceleration
tensor vanishes.

The acceleration tensor is invariant under coordinate
transformations. However it is not invariant under SO(3,1) which
means it is frame-dependent. Therefore given a set of tetrad field
the translational acceleration of the frame along $C$ follows from
$\phi_{(0)}\,^{(i)}$ and the angular velocity from $\phi_{(i)(j)}$.
Consequently the acceleration tensor is suitable to describe
geometrically an observer in space-time. It does not contain any
dynamical feature which depends on field equations.

\section{The isotropic and Homogeneous Universe}
\noindent

The cosmological principle asserts that the large-scale structure of
the Universe reveals homogeneity and isotropy~\cite{Landau}. The
general form of the line element preserving such features is the
Friedman-Robertson-Walker (FRW) line element

\begin{equation}
ds^2=-dt^2+\frac{R^2(t)}{(1+\frac{k'}{4}r^2)^2}[dr^2+r^2(d\theta^2+\sin^2\theta
d\phi^2)]\,,\label{9}
\end{equation}
where $R(t)$ is the scale factor and $k'$ is the normalized
curvature of the Universe. It assumes the values -1, 0 or 1. The
above line element is an ansatz based on cosmological postulates,
therefore the temporal dependence of scale factor will be given by
Einstein's equations. There are numerous ideas to determine the
scale factor, scalar field and vanishing curvature. In all such
cases the expansion of the Universe has to depend on the scale
factor~\cite{Frieman:2008zz}. From field equations we find out that
the scale factor obeys Friedmann's equation which reads

\begin{eqnarray}
3\left(\frac{{\dot{R}}^2+k'}{R^2}\right)-\Lambda &=& 8\pi\rho\,,\nonumber\\
\frac{2R\ddot{R}+{\dot{R}}^2+k'}{R^2}-\Lambda &=& 8\pi
p\,,\label{9.1}
\end{eqnarray}
where $\rho$ is the mean density of the Universe, p is the pressure
and $\Lambda$ is the cosmological constant.

The characterization of the field velocity $U^\mu$ of an observer is
given by the component $a=(0)$ of the tetrad $e_a\,^\mu$ by the
identification $U^\mu=e_{(0)}\,^\mu$. It is chosen using the
following tetrad field

\begin{equation}
e^{a}\,_{\mu}=\left(
                \begin{array}{cccc}
                  A & C & 0 & 0 \\
                  B\sin\theta\cos\phi & D\sin\theta\cos\phi &
                  Er\cos\theta\cos\phi & -Fr\sin\theta\sin\phi \\
                  B\sin\theta\sin\phi & D\sin\theta\sin\phi &
                  Er\cos\theta\sin\phi & Fr\sin\theta\cos\phi \\
                  B\cos\theta & D\cos\theta & -Er\sin\theta & 0 \\
                \end{array}
              \right)\,, \label{10}
\end{equation}

where $A$, $B$, $C$, $D$, $E$ and $F$ are functions of $r$ and $t$,
only. They are given by

\begin{eqnarray}
E^2&=&F^2=\frac{R^2(t)}{(1+\frac{k'}{4}r^2)^2}\,,\nonumber\\
B^2&-&A^2=-1\,,\nonumber\\
D^2&-&C^2=\frac{R^2(t)}{(1+\frac{k'}{4}r^2)^2}\,,\nonumber\\
AC&=&BD\,.\label{11}
\end{eqnarray}
The metric tensor is given as $g_{\mu\nu}=e_{a\mu}e^{a}\,_{\nu}$.
The tetrad field is adapted to a reference frame, which will be
designated by $K'$, with a radial field velocity in relation to a
reference frame at rest, $K$, with
$U^\mu=e_{(0)}\,^\mu=(A,-g^{11}C,0,0)$. Indeed there is a freedom in
the choice of the function $C$, corresponding to the radial
component of velocity for different observers.

The acceleration tensor $\phi_{ab}$ reveals some fundamental
features and details of the observer with radial velocity. The
components of $\phi_{(0)(i)}$ are

\begin{equation}
\phi_{(0)(i)}= \frac{1}{2}g^{00}g^{11}(e_{(0)0} e_{(i)1}-e_{(0)1}
e_{(i)0})( g^{00}e_{(0)0}T_{001}+g^{11}e_{(0)1}T_{101})\,.
\label{12}
\end{equation}
There are only two relevant components of the torsion tensor in
above expression, i. e.

\begin{eqnarray}
T_{001}&=&-A\partial_{0}C+B\partial_0 D \,,\nonumber \\
T_{101}&=&\frac{1}{2}\partial_0 (g_{11})+C\partial_1 A-D\partial_1
B\,.\label{13}
\end{eqnarray}
The we have

\begin{eqnarray}
\phi_{(0)(1)}&=&\frac{1}{2}(AD-BC)g^{00}g^{11}X
\sin\theta\,\cos\phi \,,\nonumber \\
\phi_{(0)(2)}&=&\frac{1}{2}(AD-BC)g^{00}g^{11}X
\sin\theta\,\sin\phi \,,\nonumber \\
\phi_{(0)(3)}&=&\frac{1}{2}(AD-BC)g^{00}g^{11}X \cos\theta \,,
\label{14}
\end{eqnarray}
where $X$ is defined by the expression

\begin{equation}
X=\frac{B}{C}(g_{11}\partial_0 B-\partial_1
D)+\frac{1}{2}g^{11}(C\partial_0 g_{11}+A\partial_1
g_{11})\,.\label{15}
\end{equation}

The acceleration is given as

\begin{equation}
{\bf a}= \phi_{(0)(1)} \hat{{\bf x}}+\phi_{(0)(2)} \hat{{\bf y}}
+\phi_{(0)(3)} \hat{{\bf z}}\,, \label{16}
\end{equation}
leading to

\begin{equation}
{\bf a}= \frac{1}{2}(AD-BC)g^{00}g^{11}X\, \hat{\bf r}\,. \label{17}
\end{equation}

The frame treated here has no angular velocity which is a desirable
feature since it is supposed to be adapted to the movement of
galaxies. It is clear that a non-rotating frame has to be described
by a tetrad field with vanishing $\phi_{(i)(j)}$.

Then $\phi_{(i)(j)}$ reduces to

\begin{equation}
\phi_{(i)(j)}=\Big[g^{00}g^{11}(e_{(i)0} e_{(j)1}-e_{(i)1}
e_{(j)0})(
g^{00}e_{(0)0}T_{001}+g^{11}e_{(0)1}T_{101})\Big]\,,\label{18}
\end{equation}
that leads to

\begin{equation}
\phi_{(i)(j)}=0\,,\label{19}
\end{equation}
once $(e_{(i)0} e_{(j)1}-e_{(i)1} e_{(j)0})$ always vanish. It is to
be noted that an observer with radial field velocity is radially
accelerated as well, which means the expansion between $K'$ and $K$
is accelerated.

\subsection{The Flat Universe Case}
\noindent

The flat Universe is set up once the parameter $\kappa$ is equal to
zero. It worths to analyze such an Universe since it appears that
the curvature of our Universe is approximately
zero~\cite{Capozziello:2008zp,Vardanyan:2009ft,Bernardis:2000gy,1992MNRAS}.
Besides it may provide new insights into the nature of the dark
energy. The Hubble's law implies that the velocity of recession of
galaxies is roughly proportional to the mean separation between
them~\cite{Dinverno,hubble}.

A co-moving observer with a galaxy has the field velocity

\begin{equation}
U^\mu=e_{(0)}\,^{\mu}=(U^0,U^1,0,0)\,,\label{20}
\end{equation}
in the radial direction. The function A is identified with $U^0$ and
C with $-g_{11}U^1$. The radial component of field velocity is fixed
by

\begin{equation}
U^1=\frac{\dot{R}}{R}\,\,r\,,\label{20.1}
\end{equation}
where $\dot{R}$ is the time derivative of scale factor (R(t)) and
$r$ is the comoving distance from the origin. This is the Hubble's
Law with Hubble's constant as
$H=\frac{\dot{R}}{R}$~\cite{1993ApJ...403...28H,Dinverno}. Such a
movement is not geodesic since it deals with relative velocities
between $K'$ and $K$. The Earth may be taken as K, which is assumed
to be stationary, and some arbitrary galaxy as K'. In this context
what is important is the relative acceleration between galaxies.
Thus a non-geodesic acceleration could be established simulating
effects of dark energy.

With this choice, the functions appearing in the tetrad field are

\begin{eqnarray}
E&=&F=R\,,\nonumber\\
C&=&-R\dot{R}r\,,\nonumber\\
D&=&R(1+{\dot{R}}^2r^2)^{1/2}\,,\nonumber\\
B&=&R^{-1}C\,,\nonumber\\
A&=&R^{-1}D\,.\label{21}
\end{eqnarray}
Substituting this in (\ref{15}), we have

\begin{equation}
X=-{\dot{R}}^2r-R\ddot{R}r-\frac{{\dot{R}}^2r}{[1+(R\dot{R}r)^2]^{1/2}}\,.\label{22}
\end{equation}

Then using (\ref{17}) the non-geodesic acceleration is

\begin{equation}
a=\frac{1}{2R}\left({\dot{R}}^2r+R\ddot{R}r+\frac{{\dot{R}}^2r}{[1+(R\dot{R}r)^2]^{1/2}}\right)\,.\label{23}
\end{equation}
In order to compare this to observational data, it should be noted
that the first two terms of the above expression assume arbitrary
values when $r$ tend to infinity while the last term remains
constant. Thus for large values of $r$ it is possible to drop the
last term, therefore an accelerated rate of expansion between $K'$
and $K$ approximately is

\begin{equation}
a\approx\frac{1}{2}(1-q)H^2d_M\,, \label{24}
\end{equation}
where $q=-R\frac{\ddot{R}}{\dot{R}^2}$ is the deceleration parameter
and $d_M=Rr$ is the proper distance which is related to the
luminosity distance by $d_l=d_M(1+z)$.

The quantity (\ref{24}) represents the acceleration of an observer
whose field velocity obeys the Hubble's Law for big separation
between K and K'. Since the galaxies approximatively expand from
each other in a similar way, it is interpreted as the acceleration
of the Universe. Old measurements point to the current value of $q$
as being $1$ which means an approximately non-expanding
Universe~\cite{Dinverno}. However new data aims to a negative value
of $q$, this is taken as an indication of the existence of dark
energy~\cite{Xu:2007nj,Gong:2006gs,Rahaman:2005sd}. Let's see the
predictions of Eq. (\ref{24}) for different values of $q$.

Cosmography is an important tool to map the expansion of the
Universe. In reference \cite{Lemetz} the deceleration parameter is
reconstructed using a combination of SN Ia, BAO and CMB data. Thus
for redshift around $z\approx 1.2$, we see $q\approx 0.2$, then
using $H\approx 71\,Km\,s^{-1}\,Mpc^{-1}$, the non-geodesic
acceleration is

$$a\approx 6.9 \,Km^2\,s^{-2}\,pc^{-1}\,$$
where the luminosity distance is approximately given by
$d_l=(c\,z/H)(1+(1-q)z/2)$ and it is used $c=3.10^8\,m/s$ rather
than the unity. For $z\approx 0.2$ we see $q\approx -0.5$ and
$a\approx 3.1\,Km^2\,s^{-2}\,pc^{-1}$. Now it is due to
experimentalists to confirm (or not) the above prediction for the
acceleration of the Universe.

The acceleration in (\ref{24}) was obtained from the acceleration
tensor given by Eq. (\ref{8}), and the latter is constructed out of
the torsion tensor, which is ultimately responsible for the
non-geodesic acceleration of the Universe. It is important to note
that expression (\ref{24}) is always positive even when $0<q<1$ and
dark energy could be associated to the non-geodesic acceleration of
the expanding Universe composed of ordinary matter. In this case the
dark energy may be an effect due to the torsion tensor. In a general
case, maybe in an Universe with $k\neq0$, an acceleration could
exist due to torsion and curvature, however in such a case the
effect of accelerated reference frames has to be taken into account
together with the presence of energy-momentum source in order to
explain the observational data. The exclusion of such a possibility
will lead to an erroneous analysis since it is impossible to express
Eq. (\ref{8}) in terms of the curvature.

\section{Conclusion}
\noindent

In this paper an observer moving along the same geodesic of a galaxy
which obeys approximately the Hubble's Law has a radial
acceleration, therefore it is interpreted as the acceleration of the
Universe. The Hubble's Law gives the movement of galaxies only
approximately, thus one can see that this acceleration, in fact,
does not represent the real picture in the Universe. However a
realistic field velocity can be chosen by an appropriate choice of
the function $C$ (a null radial velocity implies $C=0$).

The case of a flat Universe ($k=0$) is analyzed, leading to a finite
and positive radial acceleration even if $0<q<1$ (ordinary matter).
This seems to be in agreement with what our Universe looks like
since recent observations lead to an expanding Universe
approximately flat~\cite{Vardanyan:2009ft}. As pointed by
\cite{Durrer:2011gq} all indications for the existence of dark
energy comes from distance measurements for a given redshift, thus
for an Universe composed by ordinary matter the relation between the
luminosity distance and redshit could be affected by the presence of
the non-geodesic acceleration similarly to what happens with dark
energy. Hence it is concluded that the observed acceleration could
be an effect of the torsion tensor. In a general case when the
curvature is different from zero such an analysis plays an important
role in determining the nature of possible dark energy.

\bigskip

\begin{acknowledgement}
  I would like to thank Prof. Maluf (Universidade de Brasília) for
  helpful discussions, Prof. Santana (Universidade de Brasília) and
  Prof. Khanna (University of Alberta) for an English revision of the
  manuscript. I also would like to thank the anonymous referee for
  contributing by bringing to my attention so important points.
\end{acknowledgement}

\bibliographystyle{adp}
\bibliography{ref}

\end{document}